\documentclass[12pt]{article}
\usepackage{verbatim}
\usepackage{amsfonts}
\usepackage{graphics}
\usepackage{amsmath}
\usepackage{times}
\usepackage{appendix}
\usepackage{color}
\usepackage{enumerate}
\usepackage{fancyhdr,latexsym,amsmath,amsfonts,amssymb,amsbsy,amsthm,url}
\usepackage[margin=0.5in,footskip=0.25in]{geometry}
\usepackage{graphics,graphicx,epsfig}
\usepackage{breqn}
\usepackage{caption,subcaption,float,subfloat}
\usepackage{pdflscape}
\usepackage{hyperref}
\usepackage{float}
\newtheorem{theorem}{Theorem}[]

\usepackage[english]{babel}
\usepackage[dvipsnames]{xcolor}
\usepackage{tikz}

\numberwithin{equation}{section}

\fancypagestyle{usual}{
  \lhead[\thepage]{}
  \chead[{\it \shortauthors}]{\sc \shorttitle}
  \rhead[]{\thepage}
  \lfoot[]{}
  \cfoot[]{}
  \rfoot[]{}
  
}

\makeatletter

\renewcommand{\section}{
  \@startsection
  {section}% name
  {1}% level
  {0pt}% indent
  {1.1\baselineskip}% beforeskip
  {0.2\baselineskip}% afterskip
  {\sc \centering}% style
}

\renewcommand{\subsection}{
  \@startsection
  {subsection}% name
  {1}% level
  {0pt}% indent
  {1.1\baselineskip}% beforeskip
  {0.2\baselineskip}% afterskip
  {\sc \centering}% style
}

\renewcommand{\subsubsection}{
  \@startsection
  {subsubsection}% name
  {1}% level
  {0pt}% indent
  {1.1\baselineskip}% beforeskip
  {0.2\baselineskip}% afterskip
  {\sc \centering}% style
}

\makeatother

\begin{document}

\title{\large\sc Modeling premiums of non-life insurance companies in India}
\normalsize
\author{\sc{Kartik Sethi} \thanks{Department of Mathematics,
Indian Institute of Technology Guwahati, Guwahati-781039, Assam, India, e-mail: sethi170121021@iitg.ac.in}
\and \sc{Siddhartha P. Chakrabarty} \thanks{Department of Mathematics,
Indian Institute of Technology Guwahati, Guwahati-781039, Assam, India, e-mail: pratim@iitg.ac.in,
Phone: +91-361-2582606, Fax: +91-361-2582649}}
\date{}
\maketitle
\begin{abstract}

We undertake an empirical analysis for the premium data of non-life insurance companies operating in India, in the paradigm of fitting the data for the parametric distribution of Lognormal and the extreme value based distributions of Generalized Extreme Value and Generalized Pareto. The best fit to the data for ten companies considered, is obtained for the Generalized Extreme Value distribution.



\end{abstract}

\section{Introduction}
\label{Section_Introduction}

The usage of parametric models such as Gamma and log-normal distribution, to fit insurance claim sizes while quite common \cite{Klugman12}, is limited in its scope in terms of capturing the tail of the loss severity distribution. Accordingly, the modeling of extreme loss events or large claims, for non-life insurance companies is primarily driven by the approach of extreme value theory (EVT) \cite{Cebrian03}. The reconciliation of the two approaches, namely, the parametric models and the EVT models, needs to recognized in the paradigm of the former providing a reasonably good overall fit, but with its shortcomings at fitting the tail of the distribution, thereby necessitating the usage of the latter \cite{Cebrian03}. Beirlant et al. \cite{Beirlant04} extended the model in \cite{Cebrian03} by observing that, while the Generalized Pareto (GP) distribution provided a good fit in \cite{Cebrian03}, the model suffered from the shortcomings of a rather large threshold, and is valid only for a small percentage of the data. The extension presented in \cite{Beirlant04}, by way of a refinement of the theory of extreme values is observed to produce much better fit, even for low threshold, as well as all data. Cooray and Ananda \cite{Cooray05} noted that payouts made by insurance companies can be modeled by Pareto distribution when it comes to large losses, with the log-normal distribution accounting for the ``large loss-low frequency'' and the ``small loss-high frequency'' data. They sought to address the aspect of modeling the tail behavior, for both small and large losses, by introducing a composite log-normal-Pareto model. A comparative analysis of two composite models, namely, log-normal-Pareto  and  Weibull-Pareto, was carried out in \cite{Preda06}. Lee \cite{Lee12} undertook the modeling of fire loss severity (from Taiwan) using GP distribution, with the determination of the thresholds being carried out using mean excess plots and Hill plots. Several tests for goodness-of-fit were conducted and Value-at-Risk (VaR) as well as Expected Shortfall (ES) were determined.

A parametric alternative to fitting heavy tailed data came by the way of Log phase-type (LogPH) distributions \cite{Ahn12}, where a class of heavy tailed loss distribution was obtained, without the consideration of a separate model to fit the tail of the distribution, as was the case with extreme value models. In \cite{Araichi16}, the problem for capital requirements under Solvency was dealt with auto-regressive conditional amount (ACA) approach, to capture the evolution of claims for insurance companies and a new VaR is shown to efficiently estimate the capital required. The usage of ``transformed'' Generalized Extreme Value (GEV) and GP distribution in modeling extreme events is discussed in the thesis of Han \cite{Han03}. The GEV and GP distribution were linked in the extreme value index (EVI) in the thesis \cite{Henry08}. In his thesis \cite{Henry08}, Henry proposed a new tail index suitable for a data set that is partitioned, which is particularly useful in scenarios when one only has partitioned data available.

While most of the available literature focuses on modeling the distribution of losses to insurance companies, in terms of claims, especially the larger claims, there is little in the literature in terms of quantitative analysis of the dynamics of the premium received by the insurance companies. It needs to be recognized that the modeling of this aspect of the insurance business is critical, especially in the paradigm of the premiums received and by extension the capital requirements for the claims loss. Accordingly, in this paper, we will examine the distribution of the premiums received by insurance companies operating in India. The paper is organized the following manner. In Section \ref{Section_Methods}, we enumerate the methodology to be adopted, namely, the distributions (along with their maximum likelihood estimator and goodness-of-fit tests), the Peak Over Threshold methods and the Block Maximization method. This is followed by Section \ref{Section_Results}, where we discuss the approach, along with the presentation of the results. Finally, the concluding remarks are noted in Section \ref{Section_Conclusion}.

\section{Methods}
\label{Section_Methods}

In this section, we present a brief deliberation on the different approaches that will be used to ascertain whether a given data set exhibits a fat-tailed distribution. In particular, will briefly discuss the methods of exponential QQ-plot, Zipf plot and mean excess plot, in order to analyze the distribution of the tail in case of the data under consideration. In the approach of the exponential QQ-plot, used to see if the data exhibits fat-tails or not, one undertakes a comparison of the plot of its exponential quantiles with the quantiles of the exponential distributions \cite{Embrechts13}. The observance of a concave behavior in the exponential QQ-plot gives an indication of heavier tails than the exponential distribution. The Zipf plot is a log-log plot of the empirical survival function (in the context of the data under consideration) and turns out to be very useful for the characterization of fat-tailed distributions, and gives the necessary but not sufficient information about the tails \cite{Embrechts13}. Further, the Zipf plot can be used in order to  check for the Paretianity (power laws) in the data under consideration, and emanates from a simple observation that the survival function of the Pareto function, given by $\displaystyle{\bar{F}(x)=\frac{x}{x_{0}}^{-\alpha},~0 < x_{0} \leq x}$ implies $\displaystyle{\log(\bar{F}(x))=\alpha \log(x_{0})-\alpha\log(x)}$. Finally, the mean excess function of a random variable $X$ is defined as, $\displaystyle{M(u)=\mathbb{E}[X-u|X > u]}$, with 
its corresponding empirical mean excess function being given by, $\displaystyle{\widehat{M}(u)=\frac{\sum\limits_{i=1}^{n} (X_{i}-u)I_{X_{i}>u}}{\sum\limits_{i=1}^{n} I_{X_{i} > u}}}$ \cite{Wahlstrom13}. In the case of the GP distribution, it is observed that for $ X \sim GP_{\xi,\beta}$ (to be revisited later in this section),
then $\mathbb{E}[X] < \infty \Leftrightarrow \xi < 1$, and for which the empirical mean excess function is given by $\displaystyle{M(u)=\frac{\beta}{1-\xi}+\frac{\xi}{1-\xi}u}$.
We can plot the empirical mean excess function for a loss distribution and observe if the plot exhibits a linear trend, indicating the data follows a GP distribution. 

In order to implement the fitting of data, we make use of log-normal, GEV and GP distribution. Accordingly, we recall the cumulative distribution function (CDF) in Table \ref{tab:cdf}.
\begin{table}[hb]
\centering
\captionsetup{justification=centering}
\begin{tabular}{|l|l|}
\hline
Distribution & CDF \\ \hline
Lognormal & $LN_{\mu,\sigma}(x)=\int\limits_{-\infty}^{x}\frac{1}{\sqrt{2\pi}\sigma t}\exp\left(-\frac{1}{2}\left(\frac{\ln t-\mu}{\sigma}\right)^{2}\right)$\\ \hline
GEV & $GEV_{\theta}(x)=\begin{cases}\exp\left(-\left(1+{\theta}x\right)^{\frac{-1}{\theta}}\right)&,\theta\ne 0,\\\exp\left(-e^{-x}\right)&,\theta=0\\\end{cases}$ \\ \hline
GP & $GP_{\xi,\beta}(x)=\begin{cases}1-\left(1+\xi\frac{x}{\beta}\right)^{\frac{-1}{\xi}} &,\xi \ne 0,\\1-\exp\left(-\frac{x}{\beta}\right)&,\xi= 0\\\end{cases}$\\ \hline
\end{tabular}
\caption{CDF of the three distributions under consideration}
\label{tab:cdf}
\end{table}
Of these, GEV and GP come under the ambit of EVT, which is motivated by the necessity of fitting the extreme scenarios. There are two common approaches for identifying extremities in data namely, \textit{Block-Maxima} and \textit{Peak-Over-Threshold} \cite{Embrechts13}.
The block maximization method involves the division of the data set into $M$ time intervals, each of length $n$. Accordingly, if $X_{1m},X_{2m},\dots, X_{nm}$ is a sequence of i.i.d random variables from a time interval $m$, then the maximum values can be described as $M_{m}=\max\left(X_{1m}, \dots X_{nm}\right)$ for 
$ m=1,2,\dots, M $, with the minimum values being described analogously. The values of $M_{m}$ are standardized by the variance $\sigma_{m}$ and the
predicted value $\mu_{m}$ \textit{i.e.,} $\displaystyle{S_{m}:=\frac{M_{m}-\mu_{m}}{\sigma_{m}}}$ to determine a non-degenerated CDF. We now state the following theorem:
\begin{theorem}[Fisher-Tippet \cite{Embrechts13}]
If $\left[F\left(\frac{x-b_{n}}{a_{n}}\right)\right]$ has a non-degenerate limiting distribution as $n \to \infty$, for some constants $a_{n}$ and $b_{n}$ that depend on $n$, then,
\[\left[F\left(\frac{x-b_{n}}{a_{n}}\right)\right] \to G(x),\]
as $ n \to \infty $ for all values of $x$, where $G(x)$ is an extreme value distribution.
\end{theorem}
One major drawback of the Block-Maxima approach is that it is inefficient and ignores all but the maximum observation in each block. The Peak-Over-Threshold method does not suffer from this drawback. In this method, a set of i.i.d. random variables $X_{1},X_{2},\dots,X_{n}$ are considered, along with a certain threshold value $u$. Assuming that the right tail of the distribution is of interest, the values of exceedances $y_{1},y_{2},\dots,y_{n}$ are determined as $y_{i}=x_{i}-u$ for all realizations $X_{i}$  above the threshold $u$, above which, the distribution of exceedances is specified as $\displaystyle{F_{u}(x;u)=P(X=y+u|X>u)=\frac{F(y+u)-F(u)}{1-F(u)}}$. In order to determine the threshold in case of the considered data, we will make used of the Hill plot, which makes used of the Hill estimator given by, $\displaystyle{\widehat{\alpha}^{H}(k)=\frac{1}{\widehat{\xi}^H(k)}=\left(\frac{1}{k}
\sum\limits_{j=1}^{k}\frac{X_{j,n}}{X_{k+1,n}}\right)}$, where $X_{1,n} \geq X_{2,n} \dots \geq X_{k, n}$ are the order-statistics of our data. The Hill plot is given by $(k,\widehat{\alpha}^{H}(k))$ and then we choose our threshold to be that $ X_{k,n}$, where the plot is relatively stable with respect to $k$. The maximum likelihood estimate (MLE) \cite{Klugman12,Wahlstrom13}, for the three distributions, under consideration, are tabulated in Table \ref{tab:mle}, respectively.
\begin{table}[hb]
\centering
\captionsetup{justification=centering}
\begin{tabular}{|l|l|}
\hline
Distribution & MLE \\ \hline
Lognormal & $\hat{\mu}=\frac{\sum\limits_{i=1}^{n}\ln{x_{i}}}{n}$ and $\hat{\sigma}^2=\frac{\sum\limits_{i=1}^{n}\ln{(x_{i}-\hat{\mu})^2}}{n}$.\\ \hline
GEV & $\hat{\theta}=\arg\max\limits_{\theta \in \Theta} l(\theta)$ where $l(\theta)$ denotes the log-likelihood function.
\\ \hline
GP & $\hat{\xi}=\frac{1}{n}\sum\limits_{i=1}^{n}\ln{(1-\hat{\tau{x_i}})}$ and $\hat{\beta}=-\frac{\hat{\xi}}{\hat{\tau}}$.
\\ \hline
\end{tabular}
\caption{MLE of the three distributions under consideration}
\label{tab:mle}
\end{table}
Further, we make use of three goodness-of-fit tests, namely, the \textit{Kolmogorov-Smirnov} (KS) test statistic \cite{MatlabKS1,MatlabKS2}, the \textit{Chi-square Goodness-of-fit} (Chi-square) test \cite{MatlabChi} and the \textit{Anderson-Darling} (AD) test statistic \cite{MatlabAD}.

\section{Results}
\label{Section_Results}

The data on the premium received by ten non-life insurance companies operating in India was obtained from the website of Insurance Regulatory and Development Authority of India (IRDAI) Monthly Insurance Data \cite{IRDAI}. While the data was available for more non-life insurance companies, the analysis was carried out only for those select companies, for which the monthly insurance data was readily available for the period of April 2003 to December 2017. This was motivated by the advantage in terms of accuracy of results of hypothesis testing, in case of greater number of data points. While the methodology described in Section \ref{Section_Methods} was implemented for the data under consideration for all the insurance companies, some of the results presented in the discussion below is for Royal Sundaram, \textit{the choice of which is only for illustrative purpose}, since it is not feasible to include the corresponding results of all the other insurance companies, within the limited confines of the presentation of this article.

\begin{figure}[H]
	\centering
	\includegraphics[scale=0.6]{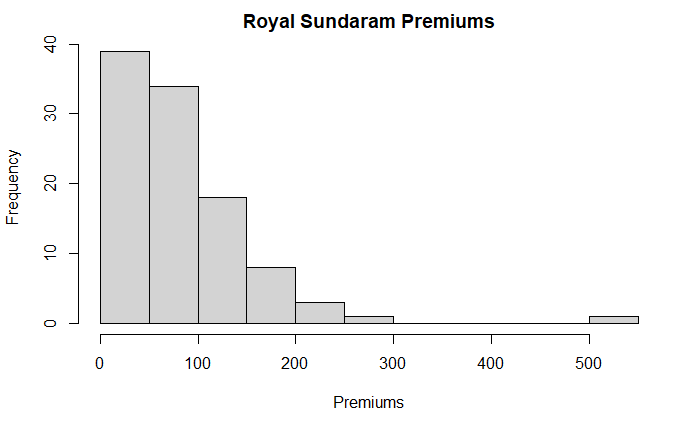}
	\caption{Histogram of Premiums for Royal Sundaram during 2003-2017}
	\label{Royal_Sundaram_Hist}
\end{figure}

\begin{figure}[H]
	\centering
	\includegraphics[scale=0.6]{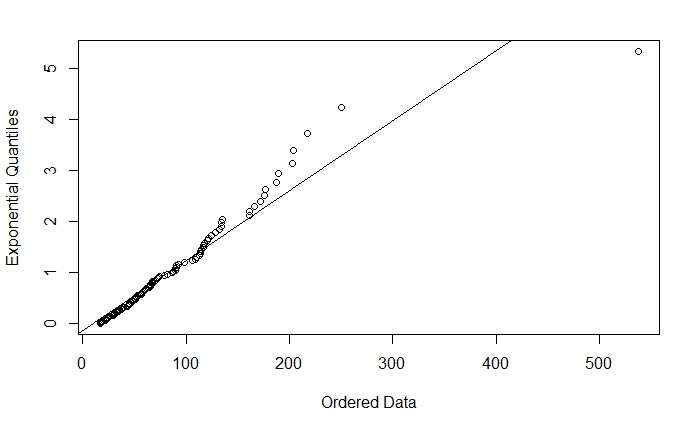}
	\caption{QQ plot of Premiums for Royal Sundaram during 2003-2017}
	\label{Royal_Sundaram_QQ}
\end{figure}

\begin{figure}[H]
	\centering
	\includegraphics[scale=0.6]{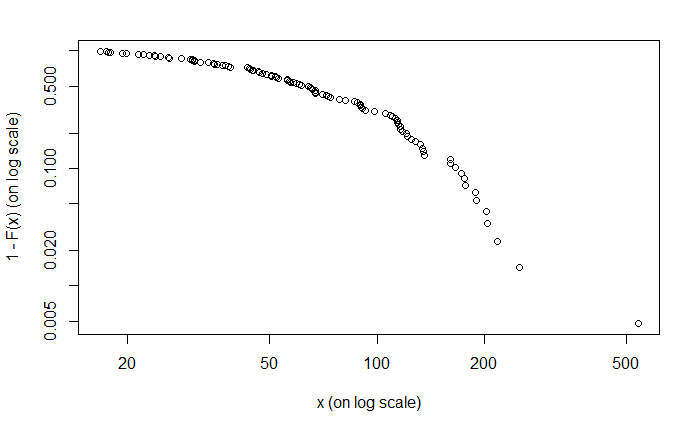}
	\caption{Zipf plot of Premiums for Royal Sundaram during 2003-2017}
	\label{Royal_Sundaram_Zipf}
\end{figure}

\begin{figure}[H]
	\centering
	\includegraphics[scale=0.6]{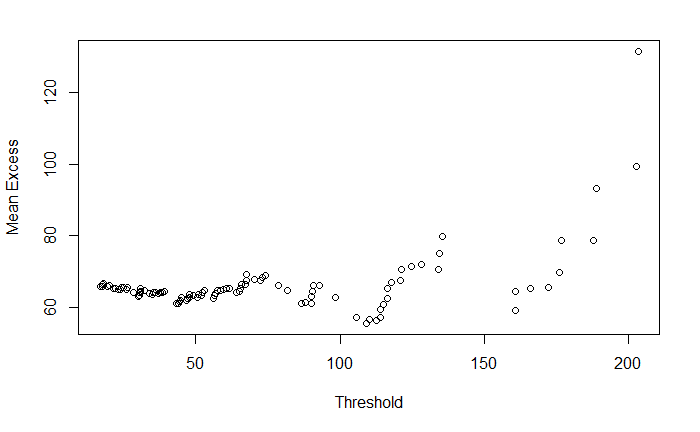}
	\caption{Mean excess plot of Premiums for Royal Sundaram during 2003-2017}
	\label{Royal_Sundaram_Meplot}
\end{figure}

The histogram of the data for Royal Sundaram is presented in Figure \ref{Royal_Sundaram_Hist}, wherein it is observed that the data is asymmetric and skewed-to-the-right or positively skewed (the mean is indeed larger than the median) and this is a indication fat-tails in corresponding data. In our QQ-plot for Royal Sundaram (see Figure \ref{Royal_Sundaram_QQ})), the initial points fall directly in line with exponential distribution's quantiles, but after that it is difficult to characterize the behaviour. The reason we are interested in the Zipf plot is because of a power law decay, which is what we are looking for in GEV Maximum Domain of Attraction. From the Zipf plot for Royal Sundaram (Figure \ref{Royal_Sundaram_Zipf}), we observe a clear negative linear slope. This is a first indication of the fat tailed nature of the data. However, as already noted, a Zipf plot verifies a necessary, but not sufficient condition. Looking at the range of the plot, the credibility of the plot seems reasonable. Given that linearity appears from the very beginning, we can observe our considered premium data as actually following a pure power law, bearing in mind that a Zipf plot is not enough. Finally, from Figure  \ref{Royal_Sundaram_Meplot}, for the mean excess plot for Royal Sundaram, we see that because of the small size of our data, it is very difficult to observe any clear trend.

\begin{figure}[H]
	\centering
	\includegraphics[scale=0.6]{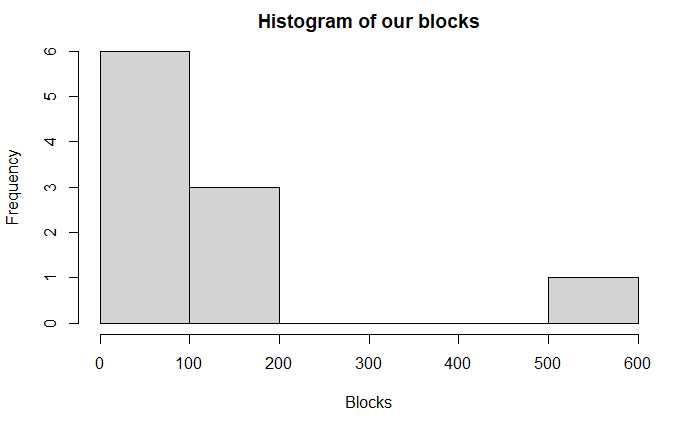}
	\caption{Histogram of the blocks for Royal Sundaram Premiums during 2003-2017}
	\label{Hist_02}
\end{figure}

\begin{figure}[H]
	\centering
	\includegraphics[scale=0.6]{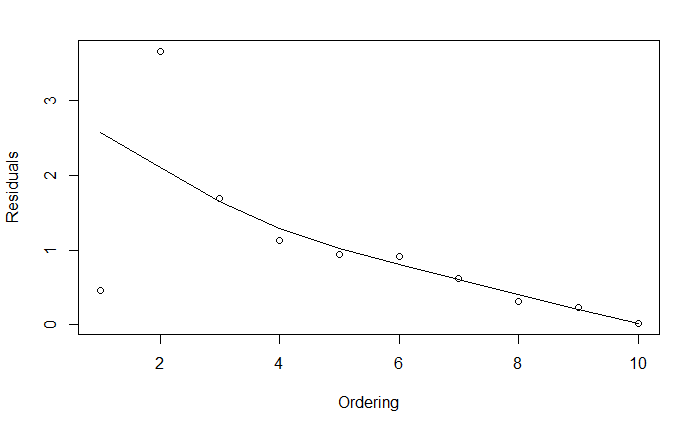}
	\caption{Residuals of Block Maxima fitting for Royal Sundaram Premiums during 2003-2017}
	\label{GEV_Residuals}
\end{figure}

\begin{figure}[H]
	\centering
	\includegraphics[scale=0.6]{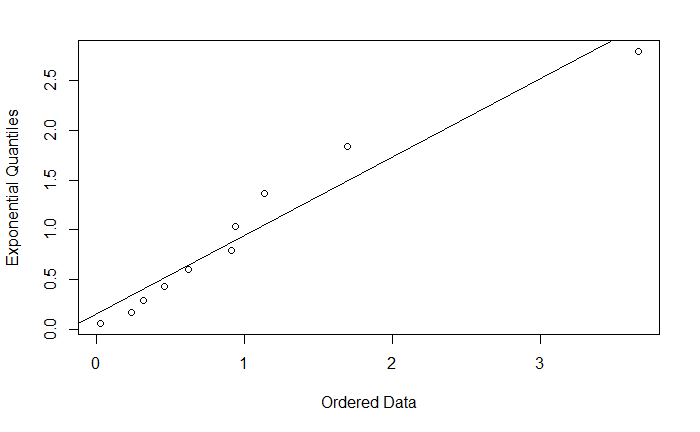}
	\caption{Exponential QQ plot of residuals of Block Maxima for Royal Sundaram Premiums during 2003-2017}
	\label{Exp_QQ}
\end{figure}

\begin{figure}[H]
	\centering
	\includegraphics[scale=0.6]{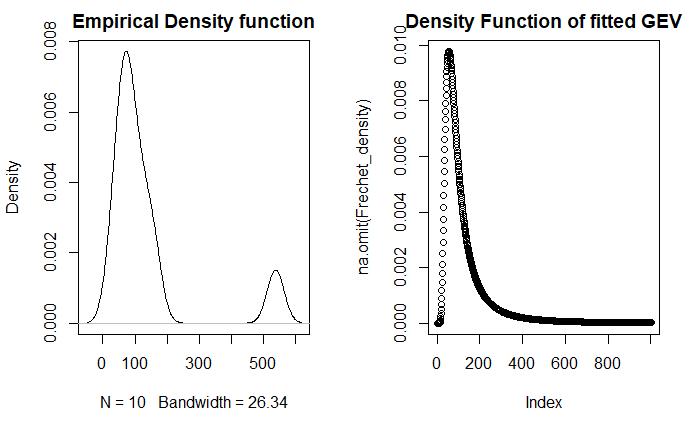}
	\caption{Actual density versus Estimated using Block Maxima approach for Royal Sundaram Premiums during 2003-2017}
	\label{Density_Comp}
\end{figure}

Because of a small data-set, it was not possible to create large number of blocks for our data, since in that case, each block would have very few data-points. We carried out the analysis by dividing the data randomly into $10$ blocks and obtained the histogram in Figure \ref{Hist_02} of our blocks.  Now we fit the GEV distribution to our data using the Block-Maxima method. We estimate the parameters using the MLE method. We can see in Figure \ref{GEV_Residuals}, that the residuals are relatively low for the tail of our data. In Figure \ref{Exp_QQ}, the empirical exponential quantiles closely follow the theoretical quantiles of our fitted GEV distribution. Finally, in Figure \ref{Density_Comp}, we compare the empirical density function and the density function of our estimated distribution.

\begin{figure}[H]
	\centering
	\includegraphics[scale=0.6]{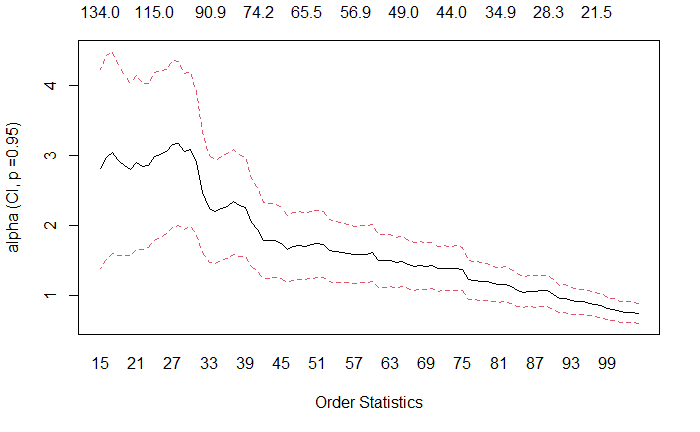}
	\caption{Hill plot of Royal Sundaram Premiums during 2003-2017}
	\label{Hill_Plot}
\end{figure}

\begin{figure}[H]
	\centering
	\includegraphics[scale=0.6]{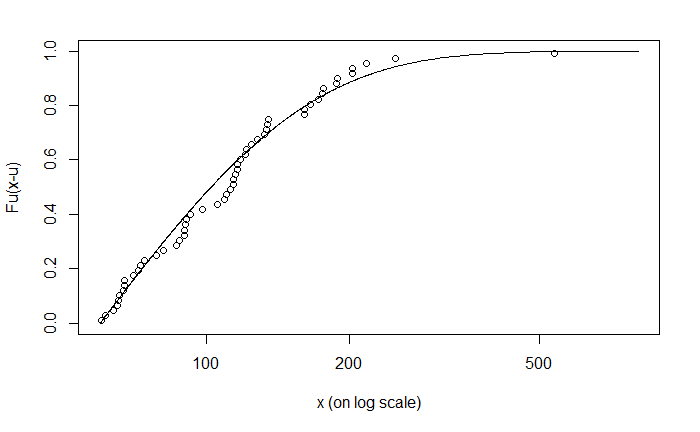}
	\caption{Peak Over Threshold: Excess Distribution}
	\label{Excess}
\end{figure}

\begin{figure}[H]
	\centering
	\includegraphics[scale=0.6]{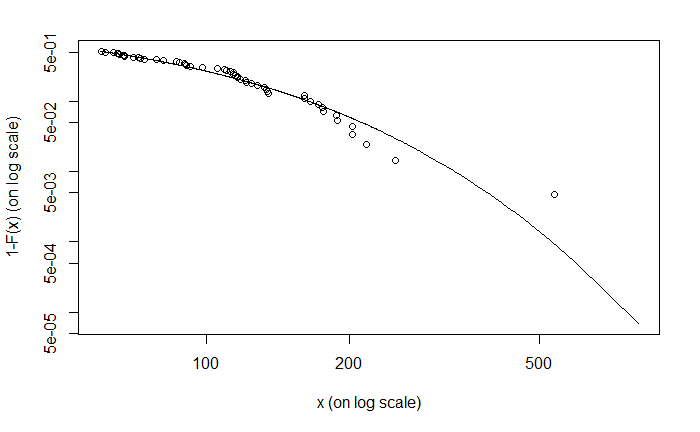}
	\caption{Peak-Over-Threshold: Tail of underlying distribution}
	\label{GPD_Tail}
\end{figure}

\begin{figure}[H]
	\centering
	\includegraphics[scale=0.8]{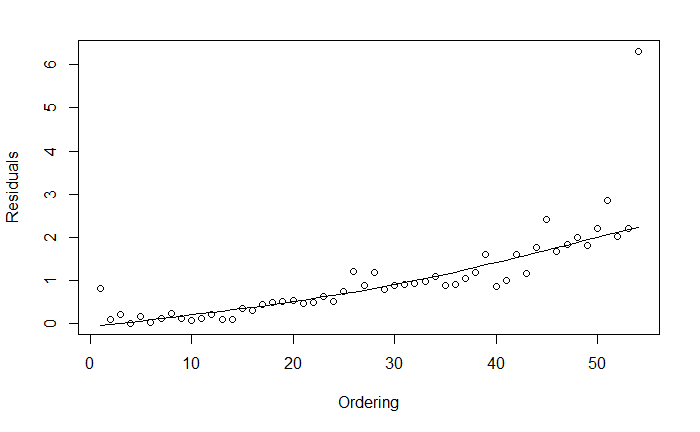}
	\caption{Peak-Over-Threshold: Residuals after fitting GPD}
	\label{GPD_Residuals}
\end{figure}

From Figure \ref{Hill_Plot} we observe that the value stabilizes around $60$, and accordingly, we take our threshold to be $60$. Now we fit the GP distribution to our data using the Peak-Over-Threshold method. We select a threshold $u$ and use the MLE, to get the best fit (see Figure \ref{Excess}). As we can see from Figure \ref{Excess}, our empirical excess distribution matches the theoretical excess distribution of the GP distribution, with the tail of the underlying distribution being presented in Figure \ref{GPD_Tail}. The residuals in Figure \ref{GPD_Residuals} of our fit shows a linear relation instead of being randomly distributed, which indicates that GPD may not be a good for the tail of our considered data.

We have tabulated the results for the log-normal, GEV and GP distributions in Table \ref{tab:lognormal}, Table \ref{tab:gev} and Table \ref{tab:gp}, respectively. We have used the Peak-Over-Threshold method for the final table of results for all the 10 insurance firms. We estimate the threshold for each firm, based on the Hill plot and then carried out the estimation of parameters. 

The results for the log-normal distribution tabulated in Table \ref{tab:lognormal}, lists the $p$-values for the three goodness-of-fit tests, namely, KS, Chi-square and the AD tests. The KS test shows that in case of six out of the ten companies, the null hypothesis results is accepted. In case of the Chi-square test, only four out of the ten companies exhibit the acceptance of the null hypothesis. Finally, in case of the AD test, the data of seven companies conform to the null hypothesis. Accordingly, given that half or less of the company data doesn't fit the null hypothesis in case of two out of three three goodness-of-fit tests, we conclude that the log-normal distribution is not an appropriate data for modeling the dynamics of evolution of insurance premium receipt data under consideration.

We next analyze the results presented in Table \ref{tab:gev}, in case of the GEV distribution. For the KS test, we observe that only in two out of the ten cases, is the null hypothesis rejected. In case of the Chi-square test, we notice that the null hypothesis is rejected in case of four companies. Again, for the AD test, only in two cases is the null hypothesis rejected. Hence it turns out that the GEV distribution provides a better fit than the log-normal distribution.

Finally, we analyze the results tabulated in Table \ref{tab:gp}. In case of the KS test, an overwhelming number of eight cases of rejection of the null hypothesis is observed. In case of the Chi-square test, only one observation results in the acceptance of the null hypothesis. Finally, for the AD test, we notice that seven out of the ten data sets, results in the rejection of the null hypothesis. Thus we conclude that the GP distribution is a worse fit to the data points in comparison to both the log-normal as well as the GEV distribution.

\section{Conclusion}
\label{Section_Conclusion}

The analysis for the considered premium data for  $10$ non-life insurance companies in India, making use of the exponential QQ-plot, Zipf plot and mean excess plot, results in a strong indication that the considered premium data for most of these companies have fat-tails. However, due to limited availability of the data, the Block Maxima method of estimating parameters turns out to be a ineffective approach, since we were unable to divide the data into sufficient number of blocks required for convergence of the maximas. Accordingly, we make use of the Peak-Over -Threshold method for our tabulated results, in order to determine the distributions that best approximate the considered data. We can see that as far as these $10$ non-life insurance companies were concerned, the GP distribution is generally not a good fit for modelling monthly insurance premium above a certain threshold. However, the GEV distribution is a good a fit for the premiums in Indian insurance industry. All hypothesis tests were done at $95\%$ confidence level and is consistent with the belief that GEV is a good fit for modelling insurance premiums in the Indian market.


\scriptsize{
	\begin{table}[h]
		\centering
		\captionsetup{justification=centering}
		\begin{tabular}{|l|l|l|l|l|l|l|l|l|}
			\hline
			Company Name     & Scale & Shape & KS       & KS P        & Chi-Square  & Chi-Square P & AD       & AD P         \\ \hline
			Royal Sundaram*  & 4.230 & 0.640 & 0.000    & 0.671       & 0.000         & 0.186      & 0.000    & 0.651        \\ \hline
			Tata-AIG         & 4.414 & 0.785 & 0.000    & 0.076       & 1.000         & 0.017      & 0.000    & 0.124        \\ \hline
			Reliance General & 4.882 & 0.642 & 1.000    & 0.003       & 1.000         & 0.000      & 1.000    & 0.029        \\ \hline
			IFFCO-Tokio      & 4.699 & 0.765 & 0.000    & 0.742       & 0.000         & 0.183      & 0.000    & 0.713        \\ \hline
			ICICI-lombard    & 5.438 & 0.907 & 1.000    & 0.031       & 1.000         & 0.012      & 0.000    & 0.137        \\ \hline
			Bajaj Allianz    & 5.155 & 0.784 & 0.000    & 0.128       & 1.000         & 0.003      & 0.000    & 0.270        \\ \hline
			HDFC CHUBB       & 4.840 & 1.082 & 0.000    & 0.440       & 0.000         & 0.238      & 0.000    & 0.383        \\ \hline
			Cholamandalam    & 4.294 & 0.802 & 0.000    & 0.643       & 0.000         & 0.086      & 0.000    & 0.421        \\ \hline
			New India*       & 5.242 & 0.886 & 1.000    & 0.001       & 1.000         & 0.000      & 1.000    & 0.016        \\ \hline
			National*        & 4.937 & 1.104 & 1.000    & 0.000       & 1.000         & 0.000      & 1.000    & 0.000        \\ \hline
		\end{tabular}
		\caption{Results for Log Normal Distribution}
		\label{tab:lognormal}
	\end{table}

	\begin{table}[h]
		\centering
		\captionsetup{justification=centering}
		\begin{tabular}{|l|l|l|l|l|l|l|l|l|l|}
			\hline
			Company Name     & Shape & Scale   & Location & KS       & KS P        & Chi-Square  & Chi-Square P & AD     & AD P   \\ \hline
			Royal Sundaram*  & 0.368 & 30.848  & 53.335   & 0.000    & 0.448       & 0.000         & 0.088      & 0.000  & 0.574  \\ \hline
			Tata-AIG         & 0.625 & 36.772  & 57.601   & 0.000    & 0.860       & 0.000         & 0.886      & 0.000  & 0.757  \\ \hline
			Reliance General & 0.008 & 66.970  & 117.728  & 0.000    & 0.085       & 1.000         & 0.000      & 0.000  & 0.151  \\ \hline
			IFFCO-Tokio      & 0.378 & 57.462  & 84.508   & 0.000    & 0.966       & 0.000         & 0.194      & 0.000  & 0.851  \\ \hline
			ICICI-lombard    & 0.264 & 153.200 & 190.516  & 0.000    & 0.270       & 0.000         & 0.065      & 0.000  & 0.323  \\ \hline
			Bajaj Allianz    & 0.285 & 98.745  & 139.234  & 0.000    & 0.206       & 1.000         & 0.008      & 0.000  & 0.347  \\ \hline
			HDFC CHUBB       & 0.706 & 82.215  & 84.998   & 0.000    & 0.357       & 0.000         & 0.253      & 0.000  & 0.288  \\ \hline
			Cholamandalam    & 0.685 & 33.423  & 49.846   & 0.000    & 0.805       & 0.000         & 0.395      & 0.000  & 0.651  \\ \hline
			New India*       & 0.198 & 131.428 & 160.477  & 1.000    & 0.004       & 1.000         & 0.000      & 1.000  & 0.016  \\ \hline
			National*        & 0.053 & 130.343 & 138.016  & 1.000    & 0.000       & 1.000         & 0.000      & 1.000  & 0.001  \\ \hline
		\end{tabular}
		\caption{Results for Generalized Extreme Value Distribution}
		\label{tab:gev}
	\end{table}
	
	\begin{table}[h]
		\centering
		\captionsetup{justification=centering}
		\begin{tabular}{|l|l|l|l|l|l|l|l|l|}
			\hline
			Company Name     & Shape  & Scale   & KS       & KS P        & Chi-Square  & Chi-Square P & AD       & AD P         \\ \hline
			Royal Sundaram*  & -0.451 & 119.918 & 1.000    & 0.002       & 1.000         & 0.005      & 1.000    & 0.002        \\ \hline
			Tata-AIG         & -0.012 & 117.953 & 1.000    & 0.000       & 1.000         & 0.000      & 1.000    & 0.002        \\ \hline
			Reliance General & -0.369 & 206.080 & 1.000    & 0.016       & 1.000         & 0.000      & 1.000    & 0.008        \\ \hline
			IFFCO-Tokio      & -0.048 & 155.470 & 1.000    & 0.014       & 1.000         & 0.016      & 1.000    & 0.048        \\ \hline
			ICICI-lombard    & -0.214 & 395.879 & 0.000    & 0.099       & 1.000         & 0.004      & 0.000    & 0.172        \\ \hline
			Bajaj Allianz    & -0.160 & 267.469 & 1.000    & 0.004       & 1.000         & 0.001      & 0.000    & 0.052        \\ \hline
			HDFC CHUBB       & 0.068  & 199.636 & 0.000    & 0.055       & 0.000         & 0.158      & 1.000    & 0.045        \\ \hline
			Cholamandalam    & -0.076 & 110.801 & 1.000    & 0.004       & 1.000         & 0.004      & 1.000    & 0.004        \\ \hline
			New India*       & -0.089 & 286.482 & 1.000    & 0.041       & 1.000         & 0.001      & 0.000    & 0.062        \\ \hline
			National*        & -0.089 & 237.886 & 1.000    & 0.001       & 1.000         & 0.000      & 1.000    & 0.000        \\ \hline
		\end{tabular}
		\caption{Results for Generalized Pareto Distribution}
		\label{tab:gp}
	\end{table}
}


\begin{thebibliography}{plain}
	
\bibitem{Klugman12}
Klugman, Stuart A., Harry H. Panjer, and Gordon E. Willmot. Loss models: from data to decisions. Vol. 715. John Wiley \& Sons, 2012.

\bibitem{Cebrian03}
Cebrian, Ana C., Michel Denuit, and Philippe Lambert. "Generalized Pareto fit to the society of actuaries’ large claims database." North American Actuarial Journal 7, no. 3 (2003): 18-36.

\bibitem{Beirlant04}
Beirlant, Jan, Elisabeth Joossens, and Johan Segers. "Generalized Pareto Fit to the Society of Actuaries Large Claims Database, Ana C. Cebrian, Michel Denuit, and Philippe Lambert, July 2003." North American Actuarial Journal 8, no. 2 (2004): 108-111.

\bibitem{Cooray05}
Cooray, Kahadawala, and Malwane MA Ananda. "Modeling actuarial data with a composite lognormal-Pareto model." Scandinavian Actuarial Journal 2005, no. 5 (2005): 321-3

\bibitem{Preda06}
Preda, Vasile, and Roxana Ciumara. "On composite models: Weibull-Pareto and Lognormal-Pareto. A comparative study." Romanian Journal of Economic Forecasting 3, no. 2 (2006).

\bibitem{Lee12}
Lee, Wo-Chiang. "Fitting the generalized Pareto distribution to commercial fire loss severity: evidence from Taiwan." The Journal of Risk 14, no. 3 (2012): 63.

\bibitem{Ahn12}
Ahn, Soohan, Joseph HT Kim, and Vaidyanathan Ramaswami. "A new class of models for heavy tailed distributions in finance and insurance risk." Insurance: Mathematics and Economics 51, no. 1 (2012): 43-52.

\bibitem{Araichi16}
Araichi, Sawssen, Christian De Peretti, and Lotfi Belkacem. "Solvency capital requirement for a temporal dependent losses in insurance." Economic Modelling 58 (2016): 588-598.

\bibitem{Han03}
Han, Zhongxian. "Actuarial modelling of extremal events using transformed generalized extreme value distributions and generalized pareto distributions." PhD diss., The Ohio State University, 2003.

\bibitem{Henry08}
Henry, John B. "Extreme value index estimation with applications to modeling extreme insurance losses and sea surface temperatures." (2008).

\bibitem{Embrechts13}
Embrechts, Paul, Claudia Klüppelberg, and Thomas Mikosch. Modelling extremal events: for insurance and finance. Vol. 33. Springer Science \& Business Media, 2013.
	
\bibitem{Wahlstrom13}
Wahlstrom, Johan. "Operational risk modeling: Theory and practice." (2013).
	
\bibitem{MatlabKS1}
\href{https://in.mathworks.com/help/stats/kstest.html}{Matlab documentation on Kolmogorov-Smirnov 1 test}
	
\bibitem{MatlabKS2}
\href{https://in.mathworks.com/help/stats/kstest2.html}{Matlab documentation on Kolmogorov-Smirnov 2 test}
	
\bibitem{MatlabChi}
\href{https://in.mathworks.com/help/stats/adtest.html}{Matlab documentation on Chi-square Goodness-of-fit test}
	
\bibitem{MatlabAD}
\href{https://in.mathworks.com/help/stats/chi2gof.html}{Matlab documentation on Anderson-Darling test}

\bibitem{IRDAI}
\href{https://www.irdai.gov.in/ADMINCMS/cms/frmGeneral_List.aspx?DF=MBFN&mid=3.2.8}{IRDAI monthly insurance data}
	
\end{thebibliography}
\end{document}